\documentclass[12pt]{article}
\usepackage{epsfig}
\usepackage{amssymb}
\usepackage{amsmath}
\usepackage{amsfonts}

\newcommand{\R}{\mathbb{R}}
\newcommand{\C}{\mathbb{C}}

\newcommand{\be}{\begin{equation}}
\newcommand{\ee}{\end{equation}}
\newcommand{\bea}{\begin{eqnarray}}
\newcommand{\eea}{\end{eqnarray}}
\newcommand{\nn}{\nonumber}
\newcommand{\kt}{\rangle}
\newcommand{\br}{\langle}

\newcommand{\ed}{\end{document}}

\newcommand{\rx}{{\rm x}}
\newcommand{\ry}{{\rm y}}

\newcommand{\rH}{{\rm H}}

\newcommand{\rh}{{\rm h}}
\newcommand{\rE}{{\rm E}}

\newcommand{\np}{\newpage}

\newcommand{\sg}{{\rm sign}}
\newcommand{\bi}{\begin{itemize}}
\newcommand{\ei}{\end{itemize}}

\begin{document}

\title{Delta-Function Potential with a Complex Coupling}
\author{\\
Ali Mostafazadeh
\\
\\
Department of Mathematics, Ko\c{c} University,\\
34450 Sariyer, Istanbul, Turkey\\ amostafazadeh@ku.edu.tr}
\date{ }
\maketitle

\begin{abstract}
We explore the Hamiltonian operator $H=-\frac{d^2}{dx^2}+
z~\delta(x)$ where $x\in\R$, $\delta(x)$ is the Dirac delta
function, and $z$ is an arbitrary complex coupling constant. For a
purely imaginary $z$, $H$ has a spectral singularity at
$E=-z^2/4\in\R^+$. For $\Re(z)<0$, $H$ has an eigenvalue at
$E=-z^2/4$. For the case that $\Re(z)>0$, $H$ has a real,
positive, continuous spectrum that is free from spectral
singularities. For this latter case, we construct an associated
biorthonormal system and use it to perform a perturbative
calculation of a positive-definite inner product that renders $H$
self-adjoint. This allows us to address the intriguing question of
the nonlocal aspects of the equivalent Hermitian Hamiltonian for
the system. In particular, we compute the energy expectation
values for various Gaussian wave packets to show that the
non-Hermiticity effect diminishes rapidly outside an effective
interaction region. \vspace{5mm}

\vspace{2mm}


\end{abstract}


\np

\section{Introduction}

The observation that a complex potential can define a consistent
unitary quantum system has recently led to a considerable research
activity. This is mostly focused on the study of the
complex-valued ${\cal PT}$-symmetric potentials $v$ for which the
Hamiltonian
    \be
    H=\frac{p^2}{2m}+v(x),
    \label{H}
    \ee
has a real discrete spectrum. In general the eigenvalue problem
for $H$ is defined along an appropriate contour $\Gamma$ in the
complex plane with suitable boundary conditions at infinity
\cite{bender-prl-1998}. This allows one to identify $H$ with a
densely defined and generally non-self-adjoint operator acting in
a separable Hilbert space ${\cal H}$, with the typical choice for
${\cal H}$ being $L^2(\Gamma)$.

The problem of whether and how one can formulate a consistent
quantum system having $H$ as its Hamiltonian has found a
satisfactory solution within the context of pseudo-Hermitian
quantum mechanics \cite{jpa-2004b,jpa-2005a,jpa-2005b,jmp-2005}.
It turns out that $H$ must be diagonalizable. In particular, there
must exist a complete basis of ${\cal H}$ consisting of
eigenvectors of $H$. This is a physical requirement of the
standard quantum measurement postulate \cite{jpa-2004b,p75}. For a
diagonalizable Hamiltonian with a discrete spectrum, the reality
of the spectrum is equivalent to the existence of a
positive-definite inner product $\br\cdot,\cdot\kt_+$ that renders
$H$ self-adjoint \cite{p23,solombrino}. The latter is a necessary
and sufficient condition for the existence of an equivalent
Hermitian Hamiltonian $h$ that acts in ${\cal H}$,
\cite{p23,jpa-2003}. These observations lead to the realization
that the physical system under investigation may be equally well
described by a Hermitian Hamiltonian within the framework of the
conventional quantum mechanics
\cite{critique,cjp-2004b,jpa-2004b}.

The above discussion also applies to complex potentials that are
not ${\cal PT}$-symmetric. The purpose of this paper is to study
in detail one of the simplest (though highly nontrivial) examples
of complex potentials that happens not to be ${\cal
PT}$-symmetric, namely the delta-function potential with a complex
coupling:
    \be
    v(x)=\zeta\delta(x),~~~~~~~\zeta\in\C.
    \label{delta}
    \ee
We wish to explore the possibility of defining a unitary quantum
system based on the standard Hamiltonian (\ref{H}) and the
potential (\ref{delta}). The reference Hilbert space
\cite{jpa-2004b} is given by ${\cal H}=L^2(\R)$ and it is not
difficult to see that the spectrum of the Hamiltonian fails to be
discrete. Because of the presence of the continuous part of the
spectrum the results reported in \cite{p1,p23} may not hold.
Nevertheless, they provide some useful guiding principles that we
will follow. Specifically, we will attempt to construct a
biorthonormal system whenever possible and use it to define an
appropriate positive-definite inner product that renders $H$
self-adjoint. An example of a successful application of this
strategy is the scattering potential \cite{jmp-2005}
    \be
    v(x)=\left\{\begin{array}{ccc}
    -i\lambda~{\rm sign}(x) &{\rm for}& |x|<\frac{L}{2}\\
    0 &{\rm for}& |x|>\frac{L}{2},\end{array}\right.,
    \label{scatter}
    \ee
where $\lambda\in\R$, $L\in\R^+$, and
    \[ {\rm sign}(x):=\left\{\begin{array}{ccc}-1&{\rm for}&x<0\\
    0&{\rm for}&x=0\\
    1&{\rm for}&x>0.\end{array}\right.\]
Note that unlike (\ref{scatter}), the delta-function potential
(\ref{delta}) fails to be ${\cal PT}$-symmetric, and as we will
see, depending on the value of $\zeta$, it may lead to the
presence of a spectral singularity \cite{naimark}.

Before starting our analysis of the properties of (\ref{delta}),
we wish to point out that complex potentials consisting of one or
more delta functions have been studied in \cite{jones-1} --
\cite{jmp-2006} and that the issue of the emergence of spectral
singularities for ${\cal PT}$-symmetric potentials has been
considered in \cite{samsonov}.

\section{Spectral Properties and Biorthonormal Systems}

Consider the time-independent Schr\"odinger equation,
    \be
    H\psi=E\psi
    \label{sch-eq}
    \ee
subject to bounded boundary conditions at $x=\pm\infty$.
Substituting (\ref{H}) and (\ref{delta}) in (\ref{sch-eq}) and
introducing the dimensionless quantities:
    \be
    \rx:=\frac{x}{\ell},~~~~z:=\frac{2m\ell\zeta}{\hbar^2},~~~~
    \rE:=\frac{2m\ell^2 E}{\hbar^2},
    \label{dim}
    \ee
where $\ell$ is an arbitrary length scale, we can express
(\ref{sch-eq}) in the form
    \be
    -\psi''(\rx)+z\delta(\rx)\psi(\rx)=\rE\psi(\rx).
    \label{sch-eq-diff}
    \ee
Clearly $\rE$ belongs to the spectrum of the dimensionless
Hamiltonian
    \be
    \rH:=\frac{2m\ell^2}{\hbar^2}\,H=
    -\frac{d^2}{d\rx^2}+z\,\delta(\rx).
    \label{dim-H}
    \ee

The solution of (\ref{sch-eq-diff}) has the form
    \be
    \psi(\rx)=\psi_k(\rx):=\left\{\begin{array}{ccc}
    A_- e^{ik\rx}+B_- e^{-ik\rx}&{\rm for}& \rx<0\\
    A_+ e^{ik\rx}+B_+ e^{-ik\rx}&{\rm for}& \rx\geq 0,
    \end{array}\right.
    \label{psi}
    \ee
where $k:=\sqrt{\rE}$, the coefficients $A_-,B_-\in\C$ are
arbitrary but not both vanishing, i.e., $|A_-|^2+|B_-|^2\neq 0$,
and
    \be
    A_+=(1-\frac{iz}{2k})A_--\frac{iz}{2k}~B_-,~~~~~~
    B_+=\frac{iz}{2k}~A_-+(1+\frac{iz}{2k})B_-.
    \label{AB}
    \ee
A straightforward implication of (\ref{psi}) and (\ref{AB}) is
that whenever $\Re(z)<0$ there is a solution $\psi_k$ with
$k=iz/2$ that belongs to $L^2(\R)$, i.e., the spectrum consists of
the obvious real, nonnegative, continuous part and a single
eigenvalue: $\rE:=-z^2/4$.

The presence of a pair $(A_-,B_-)$ of arbitrary constants in the
expression for the eigenfunctions $\psi_k$ is an indication that
the energy levels are doubly degenerate. The application of the
program of pseudo-Hermitian quantum mechanics \cite{jpa-2004b}
requires the construction of a complete biorthonormal system
consisting of the eigenfunctions of $\rH$ and $\rH^\dagger$. The
first step in this direction is to make a convenient choice for
basis eigenfunctions within each degeneracy subspace. It is most
convenient to choose one of these eigenfunctions reflectionless
\cite{reflectionless1} -- \cite{reflectionless2}. This would allow
one to express this eigenfunction using a formula that is valid
for both $\rx<0$ and $\rx\geq 0$. It is not difficult to see that
$\sin(k\rx)$ is such a reflectionless eigenfunction. Moreover, the
fact that $\sin(k\rx)$ is an odd function suggests to choose the
second basis eigenfunction to be even.\footnote{This is possible,
because the Hamiltonian is parity-invariant (${\cal
P}$-symmetric).} As we will see below, this choice simplifies the
imposition of the biorthogonality conditions considerably.

Denoting the eigenfunctions by $\tilde\psi^k_a$, with $a=1,2$
being the degeneracy label, we set
    \bea
    \tilde\psi^k_1(\rx)&:=&\frac{1}{\sqrt\pi}\, \sin(k\rx),\\
    \tilde\psi^k_2(\rx)&:=&\frac{1}{\sqrt\pi}\,\left[\cos(k\rx)+
    \frac{z}{2k}\,\sin(k\rx)\,{\rm sign}(\rx)\right].
    \eea
We can construct the following eigenfunctions of $\rH^\dagger$ by
replacing $z$ by $z^*$ in the above formulas.
    \bea
    \tilde\phi^k_1(\rx)&:=&\frac{1}{\sqrt\pi}\, \sin(k\rx),\\
    \tilde\phi^k_2(\rx)&:=&\frac{1}{\sqrt\pi}\,\left[\cos(k\rx)+
    \frac{z^*}{k}\,\sin(k\rx)\,{\rm sign}(\rx)\right].
    \eea

Next, we wish to check the validity of the biorthonormality
relation for the system $\{\tilde\psi_a^k,\tilde\phi_a^k\}$. Using
the well-known integral representation of the delta function,
$\delta(k)=(2\pi)^{-1}\int_{-\infty}^\infty e^{ikx}dx$, the parity
of the eigenfunctions, and the fact that $k+q>0$, we immediately
find
    \be
    \br\tilde\psi^k_1|\tilde\phi_1^q\kt=\delta(k-q),~~~~~~~~~~
    \br\tilde\psi^k_1|\tilde\phi_2^q\kt=
    \br\tilde\psi^k_2|\tilde\phi_1^q\kt=0.
    \label{bi1}
    \ee
Much more complicated is the derivation of\footnote{Here we used
the identity $\delta(k)=\lim_{n\to\infty}\sin(nk)/(\pi k)$.}
    \bea
    \br\tilde\psi^k_2|\tilde\phi_2^q\kt&=&
    (1+\frac{z^{*2}}{4k^2})\,\delta(k-q)+\pi z^*\delta(k)\delta(q)\nn\\
    &=&(1+\frac{z^{*2}}{4k^2})\,\delta(k-q),~~~~~~{\rm for}~~k,q>0.
    \label{bi2}
    \eea
As seen from (\ref{bi2}), the system
$\{\tilde\psi_a^k,\tilde\phi_a^k\}$ fails to be biorthogonal for
$z=\pm 2i k$, because $\tilde\phi_2^k$ is orthogonal to both
$\tilde\psi_1^q$ and $\tilde\psi_2^q$ for all $q\in\R^+$. This is
an indication of the presence of a spectral singularity, namely
$\rE=-z^2/4$, which occurs whenever $z$ is purely
imaginary\footnote{This is because $k\in\R^+$ and $z=\pm 2 i k$.}
and is consequently embedded in the continuous spectrum of $\rH$.

A spectral singularity is a serious defect that rules out the
operator as a viable candidate for a physical observable. We will
therefore only consider non-imaginary couplings $z$. In
particular, we will focus our attention on the cases where $z$
(and hence $\zeta$) has a positive real part ($\Re(z)>0$), so that
there is no eigenvalue and the spectrum is $\R^+\cup\{0\}$. In
this case, we can define
    \bea
    \psi^k_1(\rx)&:=&\frac{1}{\sqrt\pi}\, \sin(k\rx),
    ~~~~~~~~
    \psi^k_2(\rx):=\frac{\cos(k\rx)+\frac{z}{2k}\,
    \sin(k\rx)\,{\rm sign}(\rx)}{\sqrt{\pi\left(
    1+\frac{z^2}{4k^2}\right)}},
    \label{psi=}\\
    \phi^k_1(\rx)&:=&\frac{1}{\sqrt\pi}\, \sin(k\rx),
    ~~~~~~~
    \phi^k_2(\rx):=\frac{\cos(k\rx)+\frac{z^*}{2k}\,
    \sin(k\rx)\,{\rm sign}(\rx)}{\sqrt{\pi\left(
    1+\frac{z^{*2}}{4k^2}\right)}},
    \label{phi=}
    \eea
which satisfy the biorthonormality condition
    \be
    \br \psi^k_a| \phi_b^q\kt=\delta_{ab}\delta(k-q).
    \label{biortho}
    \ee
Clearly, this relation is invariant under the transformations
    \be
    \psi_a^k(\rx)\to \psi_a^{'k}(\rx):=N_a(z,k)\psi_a^k(\rx),~~~~
    \phi_a^k(\rx)\to \phi_a^{'k}(\rx):=N_a(z,k)^{-1*}\phi_a^k(\rx),
    \label{trans}
    \ee
where $N_a:\C\times\R^+\to\C$ are functions that tend to 1 as
$k\to\infty$ and $N_a(z,k)^{\pm 1}$ do not vanish except possibly
for imaginary values of $z$.

\section{Construction of a Metric Operator}

Extending the results of \cite{p1,p23} to the model under
investigation, we wish to construct a positive-definite metric
operator of the form \cite{jmp-2005,cqg-2003,ijmpa-2006}
    \be
    \eta_+=\sum_{a=1}^2\int_0^\infty dk~|\phi^k_a\kt\br\phi^k_a|.
    \label{eta=1}
    \ee
This operator defines a positive-definite inner product
$\br\cdot,\cdot\kt_+:=\br\cdot|\eta_+\cdot\kt$ that renders $H$
self-adjoint and specifies the physical Hilbert space ${\cal
H}_{\rm phys}$ of the model \cite{jpa-2004b}. Note, however, that
the metric operator (\ref{eta=1}) is not unique \cite{quasi} --
\cite{sg}; one can use $\phi^{'k}_a$ of (\ref{trans}) to construct
other admissible metric operators. Indeed, the determination of
the coefficient functions $N_a(z,k)$ that would reproduce the
usual metric operator $(\eta_+=1)$ and the ($L^2$-) inner product
in the Hermitian limit ($\Im(z)\to 0$) is a very difficult
problem. The only guiding principle is to make a simple choice for
the biorthonormal system that shares the symmetries of the
Hamiltonian. In the following we will see that the choice
(\ref{psi=})-(\ref{phi=}) made in the preceding section does
indeed fulfill this highly nontrivial requirement.

Having made a choice for $\phi_a^k$, we can try to compute the
integral kernel for $\eta_+$, namely
    \be
    \eta_+(\rx,\ry):=\br \rx|\eta_+|\ry\kt=
    \sum_{a=1}^2\int_0^\infty dk~\phi^k_a(\rx)\phi^k_a(\ry)^*.
    \label{eta=}
    \ee
Substituting (\ref{psi=}) and (\ref{phi=}) in this equation and
simplifying the result, we obtain
    \bea
    \eta_+(\rx,\ry)&=&\frac{1}{2}[\delta(\rx-\ry)-\delta(\rx+\ry)]+
    \alpha(\rx,\ry)+z\beta(\rx,\ry)\;{\rm sign}(\ry)+\nn\\
    &&z^*\beta(\ry,\rx)\;{\rm sign}(\rx)+
    |z|^2\gamma(\rx,\ry)\;{\rm sign}(\rx)\;{\rm sign}(\ry),
    \label{eta=3}
    \eea
where
    \bea
    \alpha(\rx,\ry)&:=&\frac{1}{2\pi}\int_{-\infty}^\infty dk~
    \frac{\cos(k\rx)\cos(k\ry)}{\left|1+\frac{z^2}{4k^2}\right|}
    =\frac{1}{4\pi}\,[I_0(\rx+\ry)+I_0(\rx-\ry)],
    \label{alpha}\\
    \beta(\rx,\ry)&:=&\frac{1}{4\pi}\int_{-\infty}^\infty dk~
    \frac{\cos(k\rx)\sin(k\ry)}{k\left|1+\frac{z^2}{4k^2}\right|}
    =\frac{1}{8\pi i}\,[I_1(\rx+\ry)+I_1(\ry-\rx)],
    \label{beta}\\
    \gamma(\rx,\ry)&:=&\frac{1}{8\pi}\int_{-\infty}^\infty dk~
    \frac{\sin(k\rx)\sin(k\ry)}{k^2\left|1+\frac{z^2}{4k^2}\right|}
    =\frac{1}{16\pi}\,[I_2(\rx-\ry)-I_2(\rx+\ry)],
    \label{gamma}\\
    I_n(r)&:=&\int_{-\infty}^\infty dk~\frac{e^{irk}}{k^n
    \left|1+\frac{z^2}{4k^2}\right|}=
    \int_{-\infty}^\infty dk~\frac{e^{irk}}{k^n
    \sqrt{(1+\frac{z^2}{4k^2})(1+\frac{z^{*2}}{4k^{2}})}},
    \label{In}
    \eea
$r\in\R$, and $n=0,1,2$. As seen from (\ref{eta=3}) -- (\ref{In}),
the calculation of $\eta_+(x,y)$ reduces to that of $I_n(r)$. The
latter cannot be evaluated in a closed form. We will construct a
series expansion for $I_n(r)$ that would allow for a perturbative
treatment of the problem.

First, we introduce
    \bea
    a&:=&\frac{\Re(z)^2}{4},~~~~~b:=\frac{\Im(z)^2}{4},~~~~~
    \epsilon:=\frac{\Im(z)}{\Re(z)}=\sqrt{\frac{b}{a}},
    \label{param1}\\
    s&:=&\frac{\Re(z) r}{2}=\sqrt a~r,~~~~~~~~~~~
    q:=\frac{2k}{\Re(z)}=\frac{k}{\sqrt a},
    \label{param2}
    \eea
where $\Re$ and $\Im$ stand for the real and imaginary part of
their argument, respectively.\footnote{Note that $\Re(z)$ and $a$
are both positive.}

It is not difficult to show that
    \be
    I_n(r)=\left.a^{(1-n)/2}\int_{-\infty}^\infty dq~
    \dfrac{e^{isq}\,q^{2-n} f(q,\epsilon)}{q^2+1}
    \right|_{s=\sqrt a~ r},
    \label{I1=}
    \ee
where
    \be
    f(q,\epsilon):=\left(1+\frac{[\epsilon^2+2(1-q^2)]\epsilon^2}{
    (q^2+1)^2}\right)^{-1/2}.
    \label{h=}
    \ee
In view of these equations, we may use $\epsilon$ as an
appropriate non-Hermiticity parameter. In the following we will
construct a perturbative expansion of the metric operator in terms
of $\epsilon$.

Expanding $f(q,\epsilon)$ in powers of $\epsilon$, we have
    \be
    f(q,\epsilon)=1+\frac{q^2-1}{(q^2+1)^2}~\epsilon^2+
    \frac{q^4-4q^2+1}{(q^2+1)^4}~\epsilon^4+{\cal O}(\epsilon^6),
    \label{h=expand}
    \ee
where ${\cal O}(\epsilon^N)$ stands for terms of order $N$ and
higher in powers of $\epsilon$. Substituting (\ref{h=expand}) in
(\ref{I1=}) and evaluating the resulting integrals yields, after a
very lengthy calculation that is partly done by Mathematica, the
following remarkably simple result.
    \bea
    I_0(r)&=&\left.2\pi\sqrt a~\left\{\delta(s)-\frac{e^{-|s|}}{2}+
    \left[\frac{e^{-|s|}}{8}(s^2-3|s|+1)\right]\epsilon^2\right\}
    \right|_{s=\sqrt a~ r}+{\cal
    O}(\epsilon^4),
    \label{I0=exp}\\
    I_1(r)&=&\left.i\pi~{\rm sign}(s)\;e^{-|s|}\left\{1+
    \left[\frac{1}{4}(1-|s|)|s|\right]\epsilon^2\right\}
    \right|_{s=\sqrt a~ r}+{\cal
    O}(\epsilon^4),
    \label{I1=exp}\\
     I_2(r)&=&\left.\frac{\pi}{\sqrt a}~e^{-|s|}\left\{1+
    \left[-\frac{1}{4}(s^2+|s|+1)\right]\epsilon^2\right\}
    \right|_{s=\sqrt a~ r}+{\cal
    O}(\epsilon^4).
    \label{I2=exp}
    \eea

Having obtained $I_n(r)$, we are in a position to derive an
explicit perturbative expansion for the metric operator:
    \be
    \eta_+(\rx,\ry)=
    \sum_{m=0}^{N-1}\eta_+^{(m)}(\rx,\ry)~\epsilon^m+
    {\cal O}(\epsilon^N),
    \label{perturb}
    \ee
where $N=1,2,3,\cdots$ and $\eta_+^{(m)}(\rx,\ry)$ is independent
of $\epsilon$. Inserting (\ref{I0=exp}) -- (\ref{I2=exp}) in
(\ref{alpha}) -- (\ref{gamma}), using the resulting expression to
write (\ref{eta=3}) in the form (\ref{perturb}), and employing
various properties of ``$\sg$'', particularly
    \[\sg(\rx+\ry)[\sg(\rx)+\sg(\ry)]=1+\sg(\rx)\sg(\ry)=
    2\,\theta(\rx\ry),\]
where $\theta(\rx):=[1+\sg(\rx)]/2$ is the step function, we find
after miraculous cancellations of a large number of terms
    \bea
    \eta_+^{(0)}(\rx,\ry)&=&\delta(\rx-\ry),
    \label{e-zero}\\
    \eta_+^{(1)}(\rx,\ry)&=&\frac{i\Re(z)}{4}\,\left[
    \theta(\rx\ry)\;e^{-\Re(z)|\rx-\ry|/2}+
    \theta(-\rx\ry)\;e^{-\Re(z)|\rx+\ry|/2}\right]\,\sg(\ry^2-\rx^2),
    \label{e-one}\\
    \eta_+^{(2)}(\rx,\ry)&=&\frac{\Re(z)}{16}\left\{
    \left[-\Re(z)|\rx-\ry|\,\theta(\rx\ry)+
    \theta(-\rx\ry)\right]\,e^{-\Re(z)|\rx-\ry|/2}+\right.\nn\\
    &&\hspace{1.3cm}\left.
    \left[-\Re(z)|\rx+\ry|\,\theta(-\rx\ry)+
    \theta(\rx\ry)\right]\,e^{-\Re(z)|\rx+\ry|/2}\right\},
    \label{e-two}\\
    \eta_+^{(3)}(\rx,\ry)&=&\frac{i\Re(z)^2}{32}\,\left\{
    \theta(\rx\ry)|\rx-\ry|(1-\mbox{$\frac{1}{2}$}\Re(z)|\rx-\ry|)\,
    e^{-\Re(z)|\rx-\ry|/2}+\right.\nn\\
    &&\hspace{1.3cm}\left.
    \theta(-\rx\ry)|\rx+\ry|(1-\mbox{$\frac{1}{2}$}\Re(z)|\rx+\ry|)\,
    e^{-\Re(z)|\rx+\ry|/2}\right\}\,\sg(\ry^2-\rx^2).
    \label{e-three}
    \eea
We should emphasize that according to (\ref{e-zero}) the  metric
operator $\eta_+$ tends to the identity operator in the Hermitian
limit: $\epsilon\to 0$. This is by no means a trivial expectation.
It is a consequence of our choice for the biorthonormal system.
Moreover $\eta_+^{(m)}(\rx,\ry)$ satisfy the Hermiticity
condition, $\eta_+^{(m)}(\rx,\ry)^*=\eta_+^{(m)}(\ry,\rx)$,
manifestly. A more important property of $\eta_+^{(m)}(\rx,\ry)$
is that they define bounded (integral) operators $\eta_+^{(m)}$ in
all of $L^2(\R)$. This can be established using the fact that the
integrals $\int_{-\infty}^\infty |\eta_+^{(m)}(\rx,\ry)|d\ry$ are
bounded for all $x\in\R$, \cite[\S III.2.1]{kato}. Alternatively,
we may employ the following direct proof of the boundedness of
$\eta_+^{(m)}$ for $m=0,1,2,3$, \cite{varga}. First we observe
that $|\eta_+^{(m)}(\rx,\ry)|$  viewed as a function of $\rx$ has
an upper bound, $\mu^{(m)}(\ry)$, depending on $\ry$ such that
$c:=\int_{-\infty}^\infty \mu^{(m)}(\ry)d\ry<\infty$. This implies
that for all $\psi\in L^2(\R)$,
    \bea
    \parallel \eta_+^{(m)}\psi\parallel^2&=&
    \int_{-\infty}^\infty d\rx\left|
    \int_{-\infty}^\infty d\ry~
    \eta_+^{(m)}(\rx,\ry)\psi(\ry)\right|^2\nn\\
    &\leq &\int_{-\infty}^\infty d\rx\left(
    \int_{-\infty}^\infty d\ry~
    \sqrt{\mu^{(m)}(\rx)\,\mu^{(m)}(\ry)}\;
    |\psi(\ry)|\right)^2\nn\\
    &\leq&
    \int_{-\infty}^\infty d\rx~\mu^{(m)}(\rx)~
    \int_{-\infty}^\infty d\ry ~\mu^{(m)}(\ry)~
    \int_{-\infty}^\infty
    d\rx|\psi(\ry)|^2=c^2\parallel\psi\parallel^2,
    \label{proof}
    \eea
where we have used the identities: $|\eta_+^{(m)}(\rx,\ry)|\leq
\mu^{(m)}(\ry)$,
$|\eta_+^{(m)}(\rx,\ry)|=|\eta_+^{(m)}(\ry,\rx)|\leq\mu^{(m)}(\rx)$,
and the Schwarz inequality \cite{reed-simon}. This completes the
proof of the boundedness of $\eta_+^{(m)}$ for $m=0,1,2,3$. It
implies that at least up to the third order terms in $\epsilon$,
$\eta_+$ is a bounded operator acting in all of $L^2(\R)$.

We conclude this section by noting that although the Hamiltonian
$H$ is manifestly non-${\cal PT}$-symmetric, we can follow the
approach pursued in \cite{ijmpa-2006} to define an antilinear
symmetry generator (generalized ${\cal PT}$-operator
\cite{jmp-2003}) for this Hamiltonian according to
    \be
    \tau:=\sum_{a=1}^2\int_0^\infty
    dk~(-1)^a|\psi^k_a\kt\star\br\phi^k_a|,
    \label{tau}
    \ee
where $\star$ is complex-conjugation operator defined by
$(\star\br\xi|)|\zeta\kt:=\br\zeta|\xi\kt$, for all $
\psi,\zeta\in L^2(\R)$. In view of (\ref{psi=}), (\ref{phi=}), and
(\ref{tau}),
    \[\br\rx|\tau|\ry\kt=\sum_{a=1}^2\int_0^\infty
    dk~(-1)^a\psi^k_a(\rx)\phi_a^k(\ry)=\sum_{a=1}^2\int_0^\infty
    dk~(-1)^a\psi^k_a(\rx)\psi_a^k(\ry)^*.\]
We will not attempt to obtain a more explicit expression for
$\tau$, because unlike the metric operator $\tau$ does not enter
in the calculation of the physically relevant quantities.

\section{Equivalent Hermitian Hamiltonian}

Having obtained a perturbative expansion for the metric operator
we may proceed with the calculation of the equivalent Hermitian
Hamiltonian \cite{jpa-2003,jpa-2004b,jpa-2005b,jmp-2005,other1} --
\cite{other2}:
    \be
    h:=\eta_+^{1/2} H \eta_+^{-1/2}.
    \label{h}
    \ee
Using the exponential representation of the metric operator
\cite{bbj-prd}: $\eta_+=e^{-Q}$ with $Q=\sum_{m=1}^\infty
Q_m\epsilon^m$, the pseudo-Hermiticity relation \cite{p1}:
$H^\dagger=\eta_+\,H\eta_+^{-1}$, and the fact that
$Q_1=-\eta^{(1)}$, we first calculate the equivalent Hermitian
Hamiltonian corresponding to the dimensionless Hamiltonian H. The
result is \cite{jpa-2006}
    \be
    \rh=-\frac{d^2}{d\rx^2}+\Re(z)\,\delta(\rx)+
    \rh^{(2)}\,\epsilon^2+
    {\cal O}(\epsilon^3),
    \label{h=2}
    \ee
where
    \be
    \rh^{(2)}:=\frac{i\Re(z)}{4}\,[\eta_+^{(1)},\delta(\rx)].
    \label{h2}
    \ee
In view of (\ref{e-one}) and (\ref{h2}), we can easily compute
    \bea
    \br\rx|\rh^{(2)}|\ry\kt&=&
    \frac{i\Re(z)}{4}\,[\delta(\ry)-\delta(\rx)]\,
    \eta_+^{(1)}(\rx,\ry),\nn\\
    &=&\frac{\Re(z)^2}{16}\left[\delta(\rx)\,e^{-\Re(z)|\ry|/2}+
    \delta(\ry)\,e^{-\Re(z)|\rx|/2}\right],
    \label{h2=}
    \eea
where the latter expression is to be treated in the sense of
distributions (it is valid inside an integral).

Next, we express the Hermitian Hamiltonian $h$ in terms of the
original (unscaled) physical variables and the relevant length
scale of the problem which, as we explain below, is given by
    \be
    L:=\frac{\hbar^2}{m\,\Re(\zeta)}.
    \label{scale}
    \ee
This yields
    \be
    h=\frac{p^2}{2m}+\Re(\zeta)\,\delta(x)+\Im(\zeta)^2\;h^{(2)}+
    {\cal O}(\Im(\zeta)^3),
    \label{h=phys}
    \ee
where $h^{(2)}$ is defined in terms of its integral kernel,
    \be
    h^{(2)}(x,y):=\frac{m}{8\hbar^2}
    \left[\delta(x)\,e^{-|y|/L}+
    \delta(y)\,e^{-|x|/L}\right],
    \label{h2-kernel}
    \ee
according to
    \be
    (h^{(2)}\psi)(x):=\int_{-\infty}^\infty
    h^{(2)}(x,y)\,\psi(y)\,dy=
    c_0[\psi]\,e^{-|x|/L}+
    c_1[\psi]\,\delta(x),
    \label{h2=phys}
    \ee
    \[c_0[\psi]:=\frac{m\psi(0)}{8\hbar^2},~~~~~
    c_1[\psi]:=\frac{m}{8\hbar^2}\int_{-\infty}^\infty
    e^{-|y|/L}\psi(y)\,dy.\]
Clearly $h^{(2)}$ and consequently $h$ are nonlocal operators
\cite{jpa-2004b}. Furthermore, they are both real (${\cal
T}$-symmetric) and ${\cal P}$-symmetric.\footnote{Another property
of $h^{(2)}$ is that it eliminates odd wave functions.}

To demonstrate the physical consequences of the imaginary part of
the coupling constant $\zeta$ in the original Hamiltonian
(\ref{H}) and appreciate the meaning of the length scale $L$, we
calculate the energy expectation value for a Gaussian position
wave function centered at $x=0$ and having mean momentum $\bar
p=\hbar k$ and width $\sigma$,
    \be
    \Psi(x)=(\pi\sigma^2)^{-1/4}~e^{-\frac{x^2}{2\sigma^2}+ikx}.
    \label{gaussian}
    \ee
It is important to note that we work solely in the Hermitian
representation of the quantum system where $h$ represents the
Hamiltonian of the system and $x$ the position operator.
Naturally, we view $\Psi$ as an element of ${\cal H}$ which yields
the probability density of the localization of the particle in the
physical space as $|\Psi(x)|^2$. The corresponding element of
${\cal H}_{\rm phys}$ is given by $\psi=\eta_+^{-1/2}\Psi$,
\cite{jpa-2004b}.

The energy expectation value of a particle in the state described
by the (normalized) position wave function~(\ref{gaussian}) has
the form
    \be
    \br\Psi|h|\Psi\kt=\frac{\hbar^2(\sigma^{-2}+2k^2)}{4m}+
    \frac{\Re(\zeta)}{\sqrt\pi\:\sigma}+
    \left(\frac{m\,\Omega(\sigma,k)}{2^{3/2}\hbar^2}\right)
    \Im(\zeta)^2+{\cal O}(\Im(\zeta)^3),
    \label{energy-exp}
    \ee
where
    \be
    \Omega(\sigma,k):=e^{-\frac{1}{2}(k^2-L^{-2})\sigma^2}
    \left[\cos(L^{-1}k\sigma^2)-\Re\left\{
    e^{iL^{-1}k\sigma^2}{\rm erf}
    [2^{-1/2}(L^{-1}+ik)\sigma]\right\}\right],
    \label{E=}
    \ee
and ${\rm erf}(x):=2\pi^{-1/2}\int_0^x e^{-y^2}dy$ is the error
function. The presence of the exponential factor on the right-hand
side of (\ref{E=}) suggests that the non-Hermiticity effect decays
rapidly for mean momentum values $\bar p=\hbar k$ outside the
range $[-\hbar L^{-1},\hbar L^{-1}]$. Figure~1 shows the plots of
$\Omega$ as a function of $\sigma$ for various values of $k$.
\begin{figure}
\centerline{\epsffile{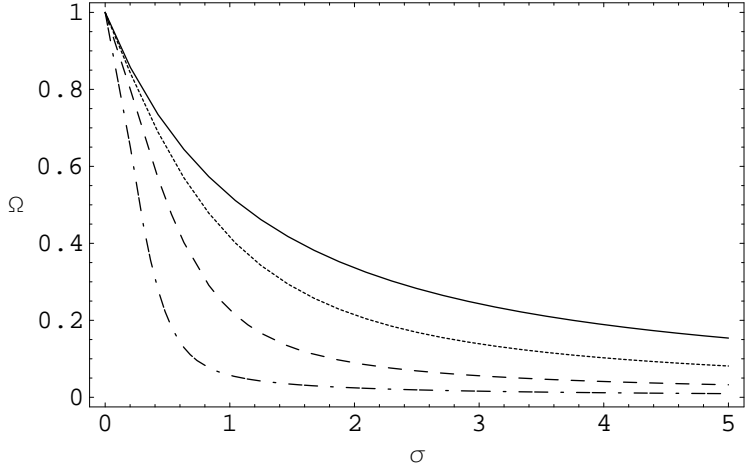}} \centerline{
\parbox{14cm}{\caption{Plots of $\Omega$ as a function of $\sigma$ for $k=0$ (the
full curve), $k=1$ (the dotted curve), $k=2$ (the dashed curve),
and $k=4$ (the dashed-dotted curve) in units where $L=1$.}}}
\label{fig1}
\end{figure}
As one increases $|k|$ the non-Hermiticity effect diminishes. The
maximum is attained for a stationary Gaussian wave packet ($k=0$)
for which
    \be
    \br\Psi|h|\Psi\kt=\frac{\hbar^2}{4m\sigma^2}+
    \frac{\Re(\zeta)}{\sqrt\pi\:\sigma}+
    \left(\frac{m\,e^{\frac{\sigma^2}{2L^2}}
    \left[1-{\rm erf}(2^{-1/2}L^{-1}\sigma)\right]}{
    2^{3/2}\hbar^2}\right)\Im(\zeta)^2
    +{\cal O}(\Im(\zeta)^3).
    \label{energy-exp-zero}
    \ee
It is not difficult to show that for such a wave packet
    {\small\[
    \br\Psi|h|\Psi\kt=\left\{\begin{array}{ccc}\!\!
    \frac{\hbar^2}{4m\sigma^2}+
    \frac{\Re(\zeta)}{\sqrt\pi\sigma}+\left(
    \frac{m L}{2\sqrt\pi\hbar^2\sigma}\right)\Im(\zeta)^2
    +\mbox{${\cal O}\left((\frac{L}{\sigma})^{3}\right)$}+
    {\cal O}(\Im(\zeta)^3)\quad\quad\quad
    \quad\quad\quad\quad\quad
    &{\rm for}&\sigma\gg L,\\ \\
    \!\!\frac{\hbar^2}{4m\sigma^2}+
    \frac{\Re(\zeta)}{\sqrt\pi\sigma}+\left(
    \frac{m}{2^{3/2}\hbar^2}\right)
    \left[1-\sqrt{\frac{2}{\pi}}\left(
    \frac{\sigma}{L}\right)+\frac{\sigma^2}{2L^2}\right]
    \Im(\zeta)^2+
    \mbox{${\cal O}\left((\frac{\sigma}{L})^{3}\right)$}
    +{\cal O}(\Im(\zeta)^3)
    &{\rm for}&\sigma\ll L.\end{array}
    \right. \]}

Next, we compute the energy expectation value for a stationary
Gaussian wave packet of width $\sigma$ and mean position $a$,
    \be
    \Psi(x)=(\pi\sigma^2)^{-1/4}~e^{-\frac{(x-a)^2}{2\sigma^2}}.
    \label{gaussian-a}
    \ee
The result is
    \be
    \br\Psi|h|\Psi\kt=\frac{\hbar^2}{4m\sigma^2}+
    \frac{e^{-\frac{a^2}{2L^2}}\Re(\zeta)}{\sqrt\pi\:\sigma}+
    \left(\frac{m\Gamma(\sigma,a)}{2^{3/2}\hbar^2}\right)\Im(\zeta)^2
    +{\cal O}(\Im(\zeta)^3),
    \label{energy-psi-a}
    \ee
where
    \[\Gamma(\sigma,a):=e^{-\frac{1}{2}\left(\frac{a^2}{\sigma^2}-
    \frac{\sigma^2}{L^2}\right)} \left\{\cosh(\mbox{$\frac{a}{L}$})-
    \mbox{$\frac{1}{2}$}e^{\frac{a}{L}}
    {\rm erf}[2^{-1/2}(\mbox{$\frac{\sigma}{L}$}+
    \mbox{$\frac{a}{\sigma}$})]-
    \mbox{$\frac{1}{2}$}e^{-\frac{a}{L}}
    {\rm erf}[2^{-1/2}(\mbox{$\frac{\sigma}{L}$}-
    \mbox{$\frac{a}{\sigma}$})]\right\}.\]
Figure~2 shows the plots of $\Gamma$ as a function of $a$ for
various values of $\sigma$. As seen from these plots the
non-Hermiticity effect is substantially smaller for mean positions
outside $[-L,L]$. Therefore, $L$ determines the range of the
non-Hermitian (nonlocal) interaction.
\begin{figure}
\centerline{\epsffile{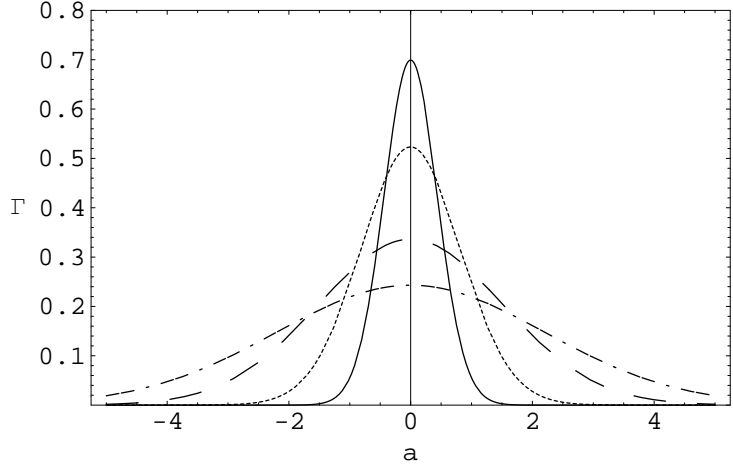}} \centerline{
\parbox{14cm}{\caption{Plots of $\Gamma$ as a function of $a$ for
$\sigma=0.5$ (the full curve), $\sigma=1$ (the dotted curve),
$\sigma=2$ (the dashed curve), and $\sigma=3$ (the dashed-dotted
curve) in units where $L=1$.}}} \label{fig2}
\end{figure}

In order to make a crude estimate for the magnitude $L$, consider
the application of the delta function potential in modelling a
point defect in a one-dimensional electron gas system. If we take
the spatial size $d$ of the defect (lattice size of the crystal)
to be of the order of 1 Angstrom and the strength of the real part
of the potential\footnote{Here we write the real part of the
potential as $d^{-1}\Re(z)\delta(d^{-1}x)$ and identify
$d^{-1}\Re(z)$ with its strength.} to be of the order of 1~ev, for
an electron (of usual mass) we find $L$ to be of the order
$10^{-10}$ Angstrom! Similarly we can obtain an order of magnitude
estimate for the strength of the non-Hermitian interaction namely
$m\Re(z)^2/(8\hbar^2 d)$. This turns out to be
$10^{8}\epsilon^2$~ev. Therefore, to ensure the validity of our
perturbative calculation of $h$, we need to take $\epsilon\ll
10^{-4}$. We also recall that the non-Hermitian interaction will
be significant, if it is stronger than the thermal effects, i.e.,
$m\Re(z)^2/(8\hbar^2 d)>kT$. At room temperature ($kT\approx
10^{-2}$~ev), this implies $\epsilon>10^{-5}$.

Finally, we would like to point out that the calculation of the
energy expectation value can be performed using the
pseudo-Hermitian representation of the system. This requires
calculation of the state vector $\psi=\eta_+^{-1/2}\Psi$
corresponding to the position wave function $\Psi$. The energy
expectation value then takes the form
$\br\psi,H\psi\kt_+=\br\psi|\eta_+H\psi\kt$. This calculation is
by no means easier to perform than the one reported above
\cite{cjp-2004b,comment-2006}. It simplifies to some extend, if
one chooses $\psi$ directly, e.g., identify $\psi(x)$ with a
Gaussian wave packet. However, note that $\psi(x)$ is void of a
direct physical meaning; it is not the position wave function for
the state it describes. In order to assign a physical meaning for
$\psi$ in terms of the position of the particle, one must compute
the corresponding position wave function \cite{jpa-2004b}, namely
$\Psi=\eta_+^{1/2}\psi$.

\section{Conclusion}

In this paper we applied the machinery of pseudo-Hermitian quantum
mechanics to explore a unitary quantum system determined by a
delta function potential with a complex coupling constant $\zeta$.
For an imaginary coupling constant there exists a spectral
singularity. For $\Re(\zeta)>0$ the spectrum is purely continuous
and one can construct a complete biorthonormal system. The double
degeneracy of the spectrum complicates the choice of a
biorthonormal system. We selected an appropriate biorthonormal
system that simplified the calculations and had a symmetric
expression for the pair of basis eigenfunctions associated with
each degeneracy subspace. We then constructed the corresponding
metric operator $\eta_+$ perturbatively and showed that it tended
to the identity operator in the non-Hermitian limit and was indeed
a bounded operator at least up to and including the third order
terms that we computed. This is quite remarkable, for there are an
infinity of other biorthonormal systems such that the
corresponding metric operator is either unbounded or fails to
yield the identity operator in the Hermitian limit.

Next, we constructed the equivalent nonlocal Hermitian Hamiltonian
$h$ for the system. The nature of the nonlocality of $h$ is quite
intriguing, because it originates from a complex delta-function
potential which is actually ultra-local! This seems to be the
reason why the non-Hermiticity effect appears in the Hermitian
Hamiltonian in the form of a short range interaction, i.e. it
decays rapidly outside the interaction region: $[-L,L]$, where
$L=\frac{\hbar^2}{m\,\Re(\zeta)}$. To establish this we calculated
the expectation value of energy for various Gaussian position wave
functions. For a non-stationary Gaussian wave packet centered at
the origin, the non-Hermitian effect reaches its maximum for mean
momenta in the range $[-\hbar L^{-1},\hbar L^{-1}]$. For a
stationary wave packet, it becomes sizable whenever the mean
position of the packet lies within the interaction region
$[-L,L]$.

The results reported above show how the methods developed for
treating systems with a discrete spectrum \cite{p23,jpa-2004b}
generalize to specific models with a continuous spectrum. Such a
generalization has previously been employed in the treatment of
the ${\cal PT}$-symmetric potential (\ref{scatter}) as reported in
\cite{jmp-2005}. The delta function potential considered in the
present paper is manifestly non-${\cal PT}$-symmetric, yet we
could successfully apply the methods of pseudo-Hermitian quantum
mechanics \cite{jpa-2004b,jpa-2005a,jpa-2005b,jpa-2006} to reveal
its physical content.

\subsection*{Acknowledgment}

During the course of this work I have benefitted from helpful
discussions with Alkan Kabak\c{c}{\scriptsize{\bf l}}o\u{g}lu and
Varga Kalantarov.

 \ed